\documentclass[prd,
12pt,
groupedaddress, preprintnumbers,nofootinbib,floatfix]{revtex4-2}
\usepackage[top=3cm, bottom=2.1cm, left=2cm, right=2cm]{geometry}
\pdfoutput=1

\usepackage{comment}

\usepackage{graphicx}
\usepackage{url}
\usepackage[bookmarks, pagebackref=false]{hyperref}
\usepackage[usenames,dvipsnames]{xcolor}
\usepackage{dcolumn}% Align table columns on decimal point
\usepackage{bm}% bold math
\usepackage{bbm}
\usepackage{amsmath,amssymb,amsfonts}
\usepackage{color}
\usepackage{hyperref}
\usepackage[Symbolsmallscale]{upgreek}
\usepackage{dsfont}
\usepackage[all]{xy}
\usepackage{pstricks}
\usepackage{dsfont}%
\usepackage{mathtools}
\usepackage{placeins}
\usepackage{enumerate}
\usepackage{cases}
\usepackage{xspace}
\usepackage{bbold}
\usepackage{cancel} % to make the barred text notation
\usepackage{slashed}
\usepackage{subfigure}
\usepackage{natbib}
\usepackage{physics}
\usepackage{multirow}

\newcommand{\be}{\begin{equation}}
	\newcommand{\ee}{\end{equation}}
\newcommand{\beq}{\begin{eqnarray}}
	\newcommand{\eeq}{\end{eqnarray}}
\newcommand{\en}[1]{\ \text{#1}}
\newcommand{\paa}[1]{\left(#1\right)}
\newcommand{\pd}{\partial}

\begin{document}

\title{Vector dark matter in supercooled Higgs portal models}

\author{Mads T. Frandsen}
\email{frandsen@cp3.sdu.dk}
\affiliation{CP3-Origins, University of Southern Denmark, Campusvej 55, 5230, Denmark}

\author{Matti Heikinheimo}
\email{matti.heikinheimo@helsinki.fi}
\affiliation{Department of Physics and Helsinki Institute of Physics, University of Helsinki,
	P.O. Box 64 FI-00014, Finland}

\author{Mattias E. Thing}
\email{thing@cp3.sdu.dk}
\affiliation{CP3-Origins, University of Southern Denmark, Campusvej 55, 5230, Denmark}

\author{Kimmo Tuominen}
\email{kimmo.i.tuominen@helsinki.fi}
\affiliation{Department of Physics and Helsinki Institute of Physics, University of Helsinki,
	P.O. Box 64 FI-00014, Finland}    

\author{Martin Rosenlyst}
\email{martin.jorgensen@physics.ox.ac.uk}
\affiliation{Rudolf Peierls Centre for Theoretical Physics, University of Oxford, 1 Keble Road, Oxford OX1
3NP, United Kingdom}

%%%%%%%%%%%%%%%%%%%%%%%%%%%%%%%%%%%%%%%%%%%%%%%
%%%%%%%%%%%%%%%%%%%%%%%%%%%%%%%%%%%%%%%%%%%%%%%
\begin{abstract}
    \noindent
We consider extensions of the Standard Model by a hidden sector consisting of a gauge field coupled with a scalar field. In the absence of dimensionful parameters in the tree-level potential, radiative symmetry breaking will induce the electroweak scale of the Standard Model and generate mass for the hidden sector gauge field. We consider both $U(1)_{\rm{D}}$ and $SU(2)_{\rm{D}}$ dark sector gauge groups and focus on probing the models with a combination of direct detection experiments and gravitational wave observatories. We find that recent dark matter direct detection results significantly constrain the parameter space of the models where they can account for the observed dark matter relic density via freeze-out. The gravitational wave signals originating from strongly first-order electroweak phase transition in these models can be probed in future gravitational wave observatories such as the Laser Interferometer Space Antenna. We show how the projected results complement direct detection experiments and can help probe parameter space near the neutrino floor of direct detection.
\end{abstract}
\preprint{HIP-2022-35/TH}
\maketitle
%%%%%%%%%%%%%%%%%%%%%%%%%%%%%%%%%%%%%%%%%%%%%%%
%%%%%%%%%%%%%%%%%%%%%%%%%%%%%%%%%%%%%%%%%%%%%%%

\section{Introduction}
    Despite the success of the Standard Model (SM) of particle physics, there are many phenomena that it does not explain and that appear to require new particles and interactions. One such phenomenon is the missing mass problem, which is inferred from observations of a wide range of astrophysical systems including galaxy clusters \cite{zwicky1933}, galaxies \cite{rubin1983}, and the cosmic microwave background radiation (CMB)~\cite{planck2020}. 
    A possible solution to the missing mass problem is cold dark matter, constituted by one or more new stable and neutral massive particles.
    However, the particle nature of dark matter (DM) remains unknown~\cite{Bergstrom:2000pn, Bertone:2016nfn, deSwart:2017heh}. 
    
    The cosmological observations of the light element abundance and CMB radiation spectrum imply that the SM degrees of freedom were in thermal equilibrium in the early Universe~\cite{Kawasaki:2000en, Hannestad:2004px, Ichikawa:2005vw, DeBernardis:2008zz,Planck:2018vyg}. 
    
    If DM was also in thermal equilibrium in the early Universe via interactions between the DM and the SM the observed relic abundance of DM today can arise from thermal decoupling during the evolution of the Universe. These DM-SM interactions also offer the prospect of detecting DM in experiments such as DM direct detection. The most studied example of this is the Weakly Interacting Massive Particle (WIMP) paradigm. However, the simplest WIMP models where DM interacts via the SM weak interactions are very strongly constrained by direct detection experiments.
    It is, therefore, important to explore different types of DM models such as hidden sectors coupled with the SM via portal interactions.

    In this paper, we analyze two simple $U(1)_{\rm{D}}$ and $SU(2)_{\rm{D}}$ hidden sector DM models with vector DM candidates coupled to the SM via a new SM singlet scalar through the Higgs portal. The models feature scale invariance of the tree-level Lagrangian and radiative electroweak symmetry breaking \cite{gildener1976}. One scalar mass eigenstate is SM Higgs-like, with a mass that can be set to the experimentally observed value of $125.46\pm0.16$ GeV \cite{higgsmass}. The other eigenstate is massless at tree level but obtains its mass via loop corrections. The framework of classically scale-invariant DM models with radiative symmetry breaking, mediated to the SM via the Higgs portal, has been explored in the literature; see e.g.~\cite{Chang:2007ki,Foot:2007as,Meissner:2006zh,Espinosa:2007qk,hambye2009,Carone:2013wla,Englert:2013gz,Gabrielli:2013hma,mohamadnejad2019, mohamadnejad2020, strumia2013,Baldes:2018emh}. 
    
    In this paper, we study how these simple $U(1)_{\rm{D}}$ and $SU(2)_{\rm{D}}$ models can be tested with a combination of direct detection and gravitational wave observations. Direct detection experiments have provided very stringent constraints on interactions of weak-scale dark matter with nuclei. Currently, the most stringent constraints come from the recent PandaX-4T and LZ (2022) experiments \cite{pandax4t, LZ2022}. The models can be parametrized by two parameters --- the DM mass and gauge coupling. We identify the viable parameter space where the models  reproduce the observed relic density $\Omega h^2= 0.120$ \cite{planck2020} and get a highly testable relation between these two parameters

    Radiative symmetry breaking in classically scale-invariant models typically results in a strongly first-order electroweak phase transition (EWPT)~\cite{Konstandin:2011dr,Marzola:2017jzl}. Such a first-order EWPT could be relevant for baryogenesis and produces gravitational wave signals which could be observable in upcoming gravitational wave experiments such as Laser Interferometer Space Antenna (LISA) \cite{lisa2016}.
    
    We present a careful examination of the first-order phase transition using different numerical packages in order to characterize the theoretical uncertainty in the predictions. For the primary result of this paper, we use the python package \texttt{CosmoTransitions} \cite{wainwright2012}. Because of the large amount of supercooling for some of the parameter space one needs to be careful about computing the transition, but also in computing the correct gravitational wave signal as Coleman-Weinberg-like models face suppression which can be mitigated by strong supercooling \cite{ellis2020}.

\section{Definitions of the models}
    We consider two models where the SM is extended with a hidden sector gauge group, $U(1)_{\rm{D}}$ and $SU(2)_{\rm{D}}$ respectively, and a new scalar field $S$ charged under the gauge group. Spontaneous symmetry breaking of the hidden sector gauge group takes place when $S$ acquires a vacuum expectation value (VEV) and leads to new massive vector DM candidates. The first model we consider is the $U(1)_{\rm{D}}$ extension defined by the Lagrangian \cite{mohamadnejad2019}
    \begin{gather}
        \mathcal{L}_{\rm{U(1)_D}} = \mathcal{L}_{\rm{SM}}^0 - \frac{1}{4} V_{\mu\nu} V^{\mu\nu} + (D_\mu S)^*(D^\mu S)-V(H,S),
    \end{gather}
    where $\mathcal{L}_{\rm{SM}}^0$ is the SM Lagrangian without the Higgs potential, $S$ is the complex scalar SM singlet and $V^\mu$ is the $U(1)_{\rm{D}}$ gauge field. The covariant derivative is \mbox{$D_\mu = \pd_\mu + i g V_\mu$} and the field strength tensor of the $U(1)_{\rm D}$ vector field is \mbox{$V_{\mu\nu} = \pd_\mu V_\nu - \pd_\nu V_\mu$}. The classically scale-invariant tree-level scalar potential is given by
    \begin{gather}\label{eq:potu1}
        V(H,S) = \frac{1}{6}\lambda_H(H^\dagger H)^2 + \frac{1}{6}\lambda_S(S^*S)^2 + 2\lambda_{H S}(H^\dagger H)(S^*S).
    \end{gather}
    In principle, a kinetic mixing term $B_{\mu\nu} V_{\mu\nu}$ with the SM hypercharge field $B_\mu$ could be present, but we assume this does not arise. For example, the mixing term can be explicitly prohibited by a $\mathbb{Z}_2$ symmetry under which \mbox{$V_\mu\rightarrow -V_\mu$} and all other fields are singlets. In the unitary gauge the scalar fields are written as
    \begin{gather}
        H = \frac{1}{\sqrt{2}}
        \begin{pmatrix}
            0\\
            v_1+h_1\\
        \end{pmatrix},  \quad
        S = \frac{1}{\sqrt{2}}(v_2+h_2),
    \end{gather}
    and upon symmetry breaking the VEVs $v_{1,2}$  become nonzero. 
    The SM gauge boson masses are determined by $v_1=246$ GeV while the DM mass is determined by the VEV $v_2$ via $M_V^2=g^2 v_2^2$. 
    
    The second model we consider is the similar $SU(2)_{\rm{D}}$ extension defined by the Lagrangian \cite{strumia2013}
    \begin{gather}
        \mathcal{L}_{\rm{SU(2)_D}} = \mathcal{L}_{\rm{SM}}^0 - \frac{1}{4} V_{\mu\nu}^i V_i^{\mu\nu} + (D_\mu S)^\dagger(D^\mu S)-V(H,S),
    \end{gather}
    where the DM candidate is now the $SU(2)_{\rm{D}}$ vector triplet $V_\mu^i$, $i=1,2,3$. The covariant derivative and the field strength tensor take the forms
    \begin{gather}
        D_\mu = \pd_\mu + i g V_\mu^i t^i, \quad 
        V_{\mu\nu}^i = \pd_\mu V_\nu^i - \pd_\nu V_\mu^i + g \epsilon^i_{jk}V_\mu^j V_\nu^k,
    \end{gather}
    where $t^i=\sigma^i/2$ are the SU(2) generators and $\sigma^i$ are the Pauli matrices. In this non-Abelian model kinetic mixing of $V_\mu^i$ with SM gauge fields is forbidden by gauge symmetry. The normalization of the scalar potential is again chosen as 
    \begin{gather}\label{eq:potsu2}
        V(H,S) = \frac{1}{6}\lambda_H(H^\dagger H)^2 + \frac{1}{6}\lambda_S(S^\dagger S)^2 + 2\lambda_{HS}(H^\dagger H)(S^\dagger S),
    \end{gather}
    where we use the same notation as in the $U(1)_{\rm{D}}$ case but the scalars are now both complex SU(2) doublets and in the unitary gauge given by
    \begin{gather}
        H = \frac{1}{\sqrt{2}}
        \begin{pmatrix}
            0\\
            v_1+h_1\\
        \end{pmatrix},  \quad
        S = \frac{1}{\sqrt{2}}
        \begin{pmatrix}
            0\\
            v_2+h_2\\
        \end{pmatrix}.
    \end{gather}
    
    In both of the above models the two neutral scalar states $h_{1,2}$ mix via the mass mixing matrix
    \begin{gather}
        M^2 =
        \begin{pmatrix}
            \frac{1}{2}\lambda_H v_1^2 + \lambda_{HS} v_2^2 & 2 \lambda_{HS} v_1 v_2 \\
            2 \lambda_{HS} v_1 v_2 & \frac{1}{2}\lambda_S v_2^2 + \lambda_{HS} v_1^2\\ 
        \end{pmatrix}
    \end{gather}
    This mass matrix has two eigenvalues, one of which is zero. In terms of the potential it can be understood as one direction being flat, corresponding to the massless eigenvalue. The field configuration perpendicular to the flat direction corresponds to the massive eigenstate. We identify the tree-level massive field $h$ with the SM Higgs, and the massless field $h_S$ is the scalon, which will obtain a nonzero mass at one-loop level. The mass eigenstates are connected to the gauge eigenstates via a mixing matrix of the form
    \begin{gather}
        \begin{pmatrix}
            h\\
            h_{S}\\
        \end{pmatrix}=
        \begin{pmatrix}
            \cos \alpha & -\sin \alpha\\
            \sin \alpha & \cos \alpha\\
        \end{pmatrix}
        \begin{pmatrix}
            h_1\\
            h_2\\
        \end{pmatrix},
    \end{gather}
    where the mixing angle $\alpha$ describes the mixing between the SM and DM sectors. This angle is restricted to small values by experiment, $\sin\alpha \lesssim 0.2$ for $M_{S}>M_h/2$ \cite{atlas:2020, CMS:2020}. 
    
    We choose the hidden sector gauge coupling $g$ and the dark matter mass $M_V$ as the input parameters. Using the minimization conditions the rest of the parameters of these models can then be written in uniform notation as
    \begin{gather}
        v_2 = \frac{c_{V} M_{V}}{g}, \quad \sin \alpha = \frac{v_1}{v},\\
        \lambda_{H} = \frac{3 M_h^2}{v_1^2} \cos^2\alpha, \quad
        \lambda_{S} = \frac{3 M_h^2}{v_2^2} \sin^2\alpha, \quad
        \lambda_{HS} = -\frac{M_h^2}{2v_1v_2}  \sin \alpha \cos\alpha, \label{modelparam}
    \end{gather}
    where $M_h$ is the SM-Higgs mass, fixed to the observed value $M_h=125.46$ GeV \cite{higgsmass}, $c_V=2$ for the $SU(2)_{\rm{D}}$ model and $c_V=1$ for the $U(1)_{\rm{D}}$ model. We have also defined $v^2 = v_1^2 + v_2^2$. 
    
    To see how the scalon, which is massless at tree level, obtains a mass we consider the loop corrections to the flat direction in the potential using the Gildener-Weinberg formalism \cite{gildener1976}. 
    The first-order loop corrections lead to an effective potential of the general form
    \begin{gather}
        V_{\rm{eff}}^1(h_{S}) = \frac{1}{64\pi^2} \sum_{k=1}^{n}g_k\Tilde{M}_k^4 \paa{\ln{\paa{\frac{\Tilde{M}_k^2}{\Lambda^2}}}-C_k},
        \label{eq:Voneloop}
    \end{gather}
    where $n$ is the number of states (including bosons and fermions), $\Tilde{M}_k$ refers to tree-level field-dependent masses described in Appendix \ref{appB}, which relate to the true mass as $\Tilde{M}_k=M_k \frac{h_S}{v}$ \cite{gildener1976, mohamadnejad2020}, $g_k$ is the degrees of freedom (with positive values for bosons and negative for fermions), $C_i = 3/2 (5/6)$, and $\Lambda$ is a renormalization scale. One can rewrite Equation (\ref{eq:Voneloop}) in terms of the true masses $M_k$ and get rid of the renormalization scale by minimization of the potential. This yields,
    \begin{gather}\label{eq:Voneloopshort}
        V_{\rm{eff}}^1(h_{S}) = B h_S^4\ln \paa{\frac{h_S^2}{v^2} - \frac{1}{2}}, \quad B = \frac{1}{64\pi^2 v^4} \sum_{k=1}^{n} g_k M_k^4.
    \end{gather}
    The scalon field is massless at tree level but obtains a mass from the loop corrections which can be seen by evaluating the second-order derivative of the effective potential in Equation (\ref{eq:Voneloopshort}) at the minimum, yielding
    \begin{gather}
        M_{S}^2 = \frac{1}{8\pi^2 v^2} \paa{g_{V}M_{V}^4 + 3M_Z^4+6M_W^4+M_h^4 -12m_t^4}, \label{scalonMass}
    \end{gather}
    where $g_V$ is the degrees of freedom for the vector boson: $g_V=9$ for the $SU(2)_{\rm{D}}$ model and $g_V=3$ for the $U(1)_{\rm{D}}$ \cite{mohamadnejad2019}. 
    Here $M_{S}$ is the scalon mass for each respective model and $M_{V}$ is the DM candidate.
    Notice that Equation~\eqref{scalonMass} relates the scalon and DM masses. In order for the scalon mass to be non-negative, this sets a lower bound for the DM masses. The bound is $M_V > 240$  GeV for the $SU(2)_{\rm{D}}$ model and $M_V > 185$ GeV for the $U(1)_{\rm{D}}$ model.
    
\section{Freeze-out relic density}
    The dark matter abundance in the models is determined via the freeze-out mechanism. While other effects like supercooling and filtering of DM can play a role in radiative symmetry breaking models such as those we study~\cite{hambye2018,Chway:2019kft,baker2020,Baldes:2021aph}, we will see that the standard freeze-out mechanism is operational throughout the parameter space of interest in this work.

    To see how the observed DM abundance $\Omega h^2 = 0.120 \pm 0.001$ \cite{planck2020} is generated via the freeze-out mechanism, we recall the basic formalism below. The present-day dark matter density is obtained from the Boltzmann equation
    \begin{gather}
        \frac{\dd n_{V}}{\dd t} + 3H n_{V} =  - \expval{\sigma_a v}\paa{n_{V}^2 - n_{V,{\rm{eq}}}^2}, \label{boltzmann}
    \end{gather}
    where $n_{V}$ is the number density of the dark matter, $H$ is the Hubble parameter and $\expval{\sigma_a v}$ is the thermally averaged annihilation cross section. The DM equilibrium number density in the broken phase is given as
    \begin{gather}\label{eq_density}
        n_{V}^{\rm{eq}}(T) = g_{V}\paa{\frac{M_{V}T}{2 \pi}}^{3/2}e^{-\frac{M_{V}}{T}}.
    \end{gather} 
    Equation~\eqref{boltzmann} can be rewritten using entropy conservation, the yield $Y_V = \frac{n_{V}}{s}$, and $x=\frac{M_{V}}{T}$ into the form
    \begin{gather}
        \frac{\dd Y_{V}}{\dd x}=  \frac{1}{3H} \frac{\dd s}{\dd x} \expval{\sigma_a v} \paa{Y_{V}^2 - Y_{V,{\rm{eq}}}^2},
    \end{gather}
    and solving this equation we obtain the present-day yield $Y_{V}^0$ that links to the abundance as 
    \begin{gather}\label{freeze-out_relic}
        \Omega h^2 = \frac{M_{V}s^0 Y_{V}^0 h^2}{\rho_0^c} \simeq 2.755\cdot 10^8 M_{V}s^0 Y_{V}^0 \rm{GeV}^{-1},
    \end{gather}
    where
    \begin{gather}
        s^0 = 2.8912\cdot 10^9\en{m}^{-3} , \quad \rho_0^c = 10.537h^2 \en{GeVm}^{-3} \en{for } H = h100\en{km/s/Mpc},
    \end{gather}
    and $h=0.678$.
    \begin{figure}
        \centering
            \subfigure[The DM relic density as a function of the mass of the $U(1)_{\rm{D}}$ vector DM candidate for different coupling constants, including the Planck collaboration result.]{
                \label{fig:relicU1}
                \includegraphics[height=6.45cm]{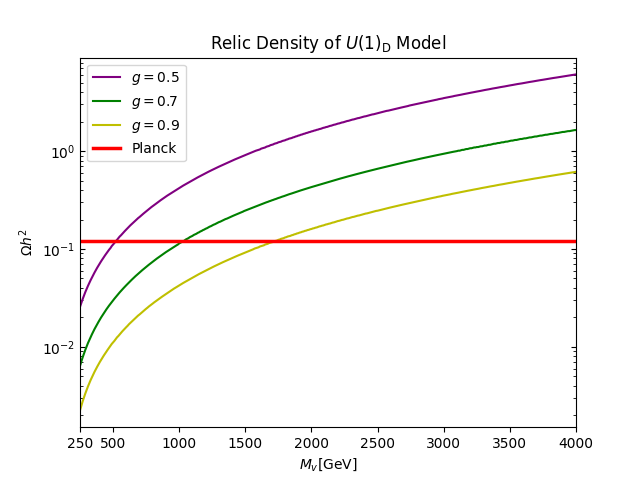}}
            \subfigure[The DM relic density as a function of the mass of the $SU(2)_{\rm{D}}$ vector DM candidate for different coupling constants, including the Planck collaboration result.]{
                \label{fig:relicSU2}
                \includegraphics[height=6.45cm]{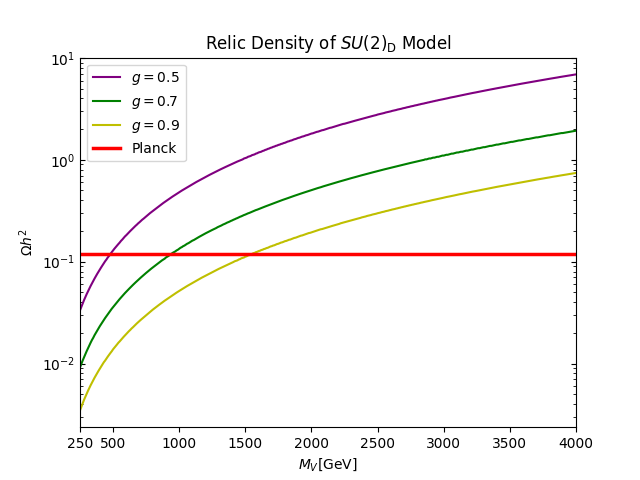}}
        \caption{The red line representing the Planck Collaboration result of $\Omega h^2 = 0.120 \pm 0.001$ is shown in red, and both models can match it via a freeze-out relic density \cite{planck2020}.}
        \label{fig:relic}
    \end{figure}
    
    To solve the Boltzmann equation numerically we use the \texttt{micrOMEGAs} package \cite{micrOMEGA}. This software uses \texttt{CalcHEP} input files with the models Feynman rules to compute the thermally averaged cross section, which we generate with the \texttt{LanHEP} package \cite{semenov2014lanhep, calchep2013}. The numerical results for the relic density for both models can be seen in Figure \ref{fig:relic}. To assess the validity of the numerical results we have compared these to the analytical result, obtained in the nonrelativistic limit and under the approximation of instantaneous freeze-out. Both of these approximations tend to overestimate the relic density. Nevertheless, the analytical result deviates only up to around 10\% for the $U(1)_{\rm{D}}$ model and slightly more for the $SU(2)_{\rm{D}}$ model, considering only the leading annihilation processes $\sigma\paa{VV \rightarrow h_S h_S}$ for the $U(1)_{\rm{D}}$ model and $\sigma\paa{V^i V^j \rightarrow h_S h_S }$ plus the semiannihilation process $\sigma\paa{V^i V^j \rightarrow V^k h_S }$  for the $SU(2)_{\rm{D}}$ model.
    
    From Figure \ref{fig:relic} it is evident that both models can reproduce the observed relic density. A larger coupling $g$ leads to more efficient annihilation of the vector DM candidate $V$ into scalons $h_S$ and, thus, the correct abundance is obtained for a correspondingly higher vector mass $M_V$. In the non-Abelian model the semiannihilation process is taken into account in the analytic approximation by defining the effective thermally averaged total annihilation cross section as
    \begin{gather}
        \expval{\sigma_a v} = \expval{\sigma_{ann} v} + \frac{1}{2} \expval{\sigma_{semi-ann} v},
    \end{gather}
     where the first term is the annihilation and the second term is the semiannihilation  cross section.
    
    The addition of the semiannihilation generally leads to more efficient annihilation, and, thus, one would expect the relic density to be lower. However, the $SU(2)_{\rm{D}}$ result in Figure \ref{fig:relicSU2} is very close to the $U(1)_{\rm{D}}$ result in Figure \ref{fig:relicU1}, which indicates that there is not much difference in the abundance for the two models considered. The origin of this is that, while the additional degrees of freedom in the non-Abelian model increase the relic density, this is balanced by the reducing effect of the semiannihilations. Concretely, the semiannihilations increase the overall thermally averaged total annihilation cross section only by roughly 15\%.
    
    Finally, we comment on the possibility of a freeze-in origin for the DM abundance in these models. In the freeze-in regime the DM particle $V$ needs to be feebly coupled to the visible sector, so that it does not reach equilibrium with the SM thermal bath in the early Universe. To achieve this, either the gauge coupling $g$ needs to be very small
    so that the vector remains decoupled while the scalar $S$ is in equilibrium, or the portal coupling $\lambda_{HS}$ can be very small, so that both the vector and the scalar remain decoupled from the SM.
    
    In the first scenario, the typical scale for the gauge coupling would be $g\sim \mathcal{O}(10^{-7})$, as seen from the approximate relation  \cite{hall2010}
    \begin{gather}
        Y_{V}(T) \sim g^2 \frac{M_{\rm{pl}}}{T},
    \end{gather}
    where $M_{\rm{pl}}$ is the reduced Planck mass. Since this process is IR dominated, the dominant production would be at the lowest kinematically allowed temperature $T \sim M_{V}$. Thus we can approximate the abundance by the replacement $T=M_{V}$ in the above to obtain
    \begin{gather}
        Y_{V}^0 \sim g^2 \frac{M_{pl}}{M_{V}}.
    \end{gather}
    Consider now the relationship between the coupling and DM mass in Equations \eqref{modelparam} and \eqref{scalonMass}. If the coupling is  $g\sim \mathcal{O}(10^{-7})$ as necessary for the freeze-in mechanism to work, the VEV $v_2$ becomes very large and the scalon mass $M_{S}$ is approximately zero. The presence of a very light scalar in the spectrum is potentially problematic, e.g. due to Higgs invisible decays, unless suppressed by a small portal coupling. On the other hand, the scenario where the portal coupling would be very small would also require a large hidden sector VEV $v_2\gg v_1$. If the gauge coupling is not very small, then this implies that the DM mass $M_V$ becomes very large. In this case, the hidden sector can be effectively populated only in the broken phase, as there is no scalar mixing in the unbroken phase. However, in this scenario there will be large supercooling, as discussed below, and the DM production should take place after reheating from thermal inflation. Now the scalar VEV is mostly in the $S$ direction $v_2\gg v_1$, so that the energy stored in the inflaton field mostly goes into $S$-quanta, but since these are feebly coupled to the SM, the reheating will be very slow and the reheating temperature suppressed. Thus, the heavy DM cannot be efficiently produced after reheating, since $T_r \ll M_V$. While there might be some way to overcome these apparent problems with freeze-in~\cite{Barman:2021lot,Barman:2022njh}, we do not consider this scenario further in this work.

\section{Inflation, reheating and supercooling}
    In the previous section, we discussed the DM abundance in the standard freeze-out scenario. 
    The situation may, however, be more complicated \cite{hambye2018, ville2021, Baldes:2021aph, baker2020}, due to a possible phase of thermal inflation characteristic of classically scale-invariant models with radiative symmetry breaking. The thermal history in the models can be summarized in terms of the following temperature thresholds:
    \begin{itemize}
        \item $T_{\rm{FO}}$: The freeze-out temperature of the DM candidate defined roughly by $n_{V}^{\rm{eq}}\expval{\sigma v}=H$.
        \item $T_{\rm{n}}$: The nucleation temperature when the probability to nucleate an expanding bubble of the broken phase vacuum inside a Hubble horizon becomes of $\mathcal{O}(1)$, approximately the temperature at which the phase transition completes.
        \item $T_{\rm{inf}}$: The temperature at the beginning of thermal inflation defined by $\rho_V=\rho_{\rm{rad}}$, where $\rho_V$ is the energy density of the false unbroken vacuum (i.e. the difference in the potential between the local minimum at $V(h_S=0)$ and the true minimum at $V(h_S=v)$) and $\rho_{\rm{rad}}$ is the energy density of the radiation-dominated Universe. When $\rho_V$ begins to dominate the energy density, inflation begins.
    \end{itemize}

    In the case of the two vector DM models discussed in this paper, the finite temperature potential includes the thermal integral summing over the bosons and fermions \cite{dolan1974},
    \begin{gather}\label{eq:firstorder}
        V_{\rm{eff}}^1(h_{S}, T) = \sum_{k=1}^{n}g_k\paa{\frac{1}{64\pi^2} \Tilde{M}_k^4 \paa{\ln{\paa{\frac{\Tilde{M}_k^2}{\Lambda^2}}}-C_k} + \frac{T^4}{2\pi^2} \int_{0}^{\infty} y^2 \ln{\paa{1 \mp e^{-\sqrt{y^2 \pm \Tilde{M}_k^2/T^2}}}}\dd y}.
    \end{gather}
    For some models, it might be necessary to consider the additional ring diagrams for the bosons, but for this investigation they can be ignored as they are insignificant \cite{carrington1992}. This thermal potential is not amenable to an analytic solution, but can be approximated using modified Bessel functions of the second kind \cite{mohamadnejad2020}. We compute the freeze-out temperature $T_{\rm{FO}}$ numerically with \texttt{micrOMEGAs}, and the nucleation temperature $T_{\rm{n}}$ numerically using \texttt{CosmoTransitions} and \texttt{Bubbleprofiler} (for cross-checking, see Appendix \ref{appC}) \cite{micrOMEGA, wainwright2012, athron2019}.

    Let us now consider the thermal history of the model depending on the order of the above three temperature thresholds. If $T_{\rm{n}} > T_{\rm{FO}}$, the phase transition completes before DM freeze-out, and the freeze-out then takes place as usual in the broken phase. This means that we can calculate the relic abundance as presented in the previous section.
    
    In the opposite case $T_{\rm{n}} < T_{\rm{FO}}$ there are three scenarios to consider. The filtered DM scenario takes place for the ordering $T_{\rm{FO}} > T_{\rm{n}} > T_{\rm{inf}}$. In this situation, there is no thermal inflation, as the phase transition completes before inflation would begin, but the DM annihilations are immediately out of equilibrium after the phase transition, and, therefore, the abundance is set by the number of DM particles that are able to enter the boundary to the broken phase, as described in \cite{baker2020}.
    
     The supercool DM scenario \cite{hambye2018}, takes place for $T_{\rm{FO}} > T_{\rm{inf}} > T_{\rm{n}}$. In this situation, there is a period of thermal inflation, which ends at $T_{\rm{n}}$. After inflation, the latent heat stored in the false vacuum is released to reheat the Universe back to temperature $T_{\rm{inf}}$, under the assumption of instant reheating, or to a lower reheating temperature for delayed reheating. However, since $T_{\rm{FO}} > T_{\rm{inf}}$, no DM is produced in reheating and the abundance is set by the amount that was present before inflation, diluted by the expansion of the scale factor and by the filtering effect as in the above scenario.
     
     Finally, there is the case where $T_{\rm{inf}} > T_{\rm{FO}}$ . In this situation, assuming instant reheating, the reheating will bring DM back to equilibrium, and the relic abundance is again obtained via the usual freeze-out mechanism as presented in the previous section.
     
     The inflation temperature is obtained by solving for $T_{\rm{inf}}$ from
    \begin{gather}
         \Delta V(T_{\rm{inf}}) =  V_{\rm{eff}}^{\rm{high}}(h_{S}, T_{\rm{inf}}) -V_{\rm{eff}}^{\rm{low}}(0, T_{\rm{inf}}) = \frac{g_* \pi^2}{30} T_{\rm{inf}}^4,
    \end{gather}
    where $V_{\rm{eff}}^{\rm{high}}(h_{S}, T)$ is the true vacuum and $V_{\rm{eff}}^{\rm{low}}(0, T)$ is the false vacuum. We find that throughout the parameter space of interest in this work, we are either in the first or the last situation described above, and the DM abundance is thus obtained via the usual freeze-out mechanism in both cases. See Appendix \ref{appA} for more on the reheating.

\section{Direct detection}
    In this section, we present the direct detection constraints on the two models. We will see that the recent results from the LZ experiment significantly affect the $SU(2)_{\rm{D}}$ model and that the $U(1)_{\rm{D}}$ model is already very constrained.

\begin{figure}
        \centering
            \subfigure[Constraints for the the $U(1)_{\rm{D}}$ model.]{
                \label{fig:constraintsU1}
                \includegraphics[height=6.45cm]{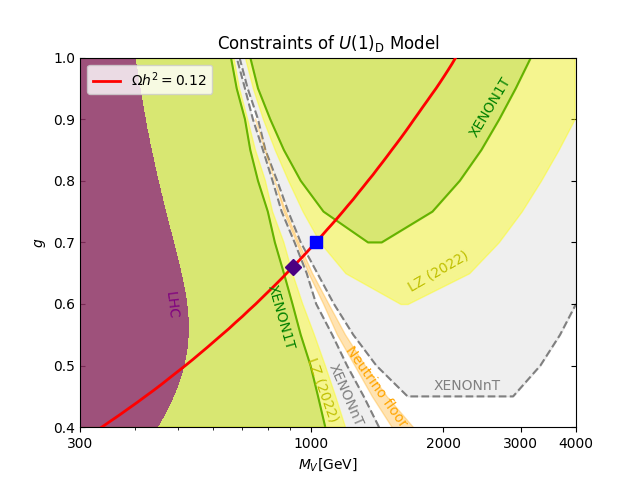}}
            \subfigure[Constraints for the the $SU(2)_{\rm{D}}$ model.]{
                \label{fig:constraintsSU2}
                \includegraphics[height=6.45cm]{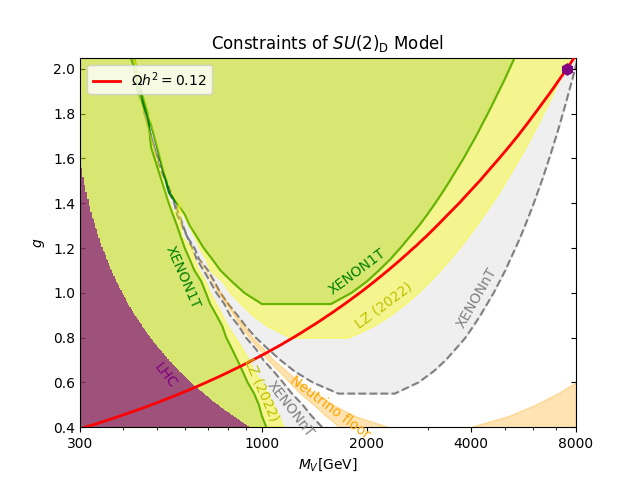}}
        \caption{The red line shows the correct relic abundance, $\Omega h^2=0.12$ \cite{planck2020}. The yellow region is excluded by the LZ (2022) experiment \cite{LZ2022}, the green region is the XENON1T experiment \cite{xenon2017}, the purple region is the LHC constraint for exotic Higgs decay, the orange region is the neutrino floor, and the gray region is the projected 90\% CL exclusion limit constraint from the XENONnT experiment \cite{xenon2020}.}
        \label{fig:constraintCombined}
    \end{figure}

    To compute the direct detection cross section, we again use the \texttt{micrOMEGAs} package \cite{micrOMEGA}. The DM coupling to nucleons arises from the scalar mixing and is mediated via exchange of the SM-like Higgs $h$ and the scalon $h_S$ in the $t$ channel leading to a spin-independent cross section with negligible difference between protons and neutrons. 
    The results of this computation for both models are shown in Figure \ref{fig:constraintCombined}. The correct relic abundance is obtained along the red solid line.

    The purple region is excluded by LHC constraints on to Higgs decays into two scalons $h\to h_S h_S$ ~\cite{ATLAS:2022yvh,CMS:2022qva}.  This process becomes kinetically forbidden for larger DM mass $M_V$, as larger DM mass leads to larger scalon mass as shown in Equation~\eqref{scalonMass}. In the orange region, the DM-nucleon cross section is below the neutrino floor, and the yellow regions indicate the exclusion limit due to the LZ experiment \cite{LZ2022}, providing a significant improvement over the XENON1T experiment shown in green \cite{xenon2017}. Finally, the gray region shows the projected exclusion limit from XENONnT \cite{xenon2020}.

    For both models we can observe a narrow wedge in the parameter space where the direct detection cross section is strongly suppressed, reaching values below the neutrino floor. This is due to a destructive interference between the $t$-channel diagrams mediated by the two scalar mass eigenstates, when the masses are degenerate \cite{Gross:2017dan,Alanne:2018zjm}.
    In the $U(1)_{\rm{D}}$ model this interference region occurs for DM mass around 0.9-1 TeV and coupling $0.65 \le g \le 0.7$. Outside of the interference region, the model cannot produce an $\mathcal{O}(1)$ fraction of DM without being excluded by direct detection, unless the DM mass is well above 10 TeV. 
    
    For the $SU(2)_{\rm{D}}$ model the new constraints from LZ alter the picture compared to the situation with the previous XENON1T limits: The relic abundance line above the interference region is now excluded for DM masses below 7.5 TeV, while prior to the LZ result there were no constraints beyond 1 TeV. In the interference region, we find the nucleation temperature for the phase transition below the QCD scale. This alters the computation for the gravitational wave signal, as the phase transition will be completed in conjunction with the QCD phase transition, as discussed in~\cite{Iso:2017uuu,vonHarling:2017yew}.  
	This picture slightly changes when including additional scalar self-energy corrections for the $SU(2)$ model~ \cite{kirkla2022}. First, the scalon mass is slightly larger than in our leading-order analysis, pushing the interference region in Figure \ref{fig:constraintsSU2} to the right. Additionally, the correction appears to slightly increase the nucleation temperature compared to our results. However, we find that overall the resulting gravitational wave (GW) signal is not significantly affected, and the GW signal prediction remains comparable to our results presented in the next section.
    
    For DM mass above 7.5 TeV the model is again allowed by direct detection. In Figure~\ref{fig:constraintCombined} we have marked three benchmark points allowed by direct detection with the blue, indigo, and purple markers. These points will be used as examples for analyzing the GW signals in the next section.
    
\section{Gravitational waves}
    The strongly first-order phase transition possible in classically scale-invariant models is interesting due to the implications for baryogenesis \cite{sakharov1967}, and due to potentially observable GW signal.
    
    To explore the gravitational wave signals, we consider the finite temperature potential in Equation~\eqref{eq:firstorder}. This potential contains a barrier between the unbroken false vacuum and the broken phase minimum, leading to a first-order phase transition. At the nucleation temperature $T_{\rm{n}}$, the phase transition will complete via the formation of bubbles of the true vacuum. The expanding and colliding bubbles deposit energy in the surrounding plasma, generating gravitational waves as described in \cite{huber2008, apreda2002, kamionkowski1994}.
    
    For the purpose of solving Equation~ \eqref{eq:firstorder} and obtaining the parameters that describe the gravitational wave signal, we use the Python package \texttt{CosmoTransitions}\cite{wainwright2012}, with custom modifications including a method of computing the $\beta$ value.  The relevant parameters are the latent heat normalized with respect to the radiation energy, $\alpha$, the inverse duration of the phase transition, $\beta$, and the nucleation temperature, $T_{\rm{n}}$,  defined as \cite{strumia2013,ellis2019},
    \begin{gather} \label{eq:gwparams}
        \alpha \equiv \eval{\frac{1}{\rho} \paa{\Delta V - \frac{T}{4}\frac{\dd \Delta V}{\dd T}}}_{T_{\rm{n}}}, \quad
        \frac{\beta}{H} \equiv T\eval{\frac{\dd(S_3/T)}{\dd T}}_{T_{\rm{n}}},
    \end{gather}
    where,
    \begin{gather}
        \Delta V =  V_{\rm{eff}}^{\rm{high}}(h_{S}, T) -V_{\rm{eff}}^{\rm{low}}(h_{S}, T), \quad
        \rho = \frac{g_e \pi^2}{30} T_{\rm{n}}^4,
    \end{gather}
    where the $g_e \approx 103$ is the number of effective degrees of freedom during the nucleation at the temperature $T_n$. Finally the Euclidean action is defined as,
    \begin{gather}
        S_3 = 4 \pi \int_{0}^{\infty} r^2 \paa{\frac{1}{2}\paa{\frac{\dd h_{S}}{\dd r}}^2 + V_{\rm{eff}}(h_{S})} \dd r,
    \end{gather}
    where $r$ is the radial distance from the center of the true vacuum bubble. 

    In order to assess the reliability of the results, we make use of two different numerical tools for computing the nucleation temperature and the $\beta$ and $\alpha$ parameters. The parameters $\alpha$ and $\beta$ depend heavily on the nucleation temperature, $T_{\rm{n}}$, so that possible errors on $T_{\rm{n}}$ will propagate to $\alpha$ and $\beta$. For the computation we use \texttt{CosmoTransitions} and \texttt{BubbleProfiler}\cite{wainwright2012, athron2019}. As shown in Appendix \ref{appD}, we obtain a smaller numerical error with \texttt{CosmoTransitions}, but the results of both numerical computations agree within uncertainty. In general, we find that for sub-TeV DM masses the nucleation temperature in the \texttt{BubbleProfiler} implementation tends to be smaller than in \texttt{CosmoTransitions}.
        
    In Figure \ref{fig:constraintCombined}, we identify three benchmark points allowed by all constraints. These benchmark points are shown in \ref{tab:GWpoints} corresponding to the indigo diamond, blue square and purple hexagon shown in Figure \ref{fig:constraintCombined}.
    \begin{figure}
        \begin{center}
            \begin{tabular}{ |c|c|c|c|c|c| }
                \hline
                Model & Benchmark point & Parameter & \texttt{CosmoTransitions} &\texttt{BubbleProfiler} & Other parameters\\
                \hline
                \multirow{3}{3.5em}{$U(1)_{\rm{D}}$} &\multirow{3}{8em}{$g=0.66$ $M_V=911\en{GeV}$ $T_c=303\en{GeV}$} & $\alpha$ & 20740 & 92180 & $S_3=986$ GeV \\
                & & $\beta/H$ & 23.8  & 39.2 & $\Delta V=1.73\cdot 10^9$ GeV \\ 
                & & $T_{\rm{n}}$ & 7.04 GeV & 4.78 GeV & $v=1402$ GeV \\ 
                \hline
                \multirow{3}{3.5em}{$U(1)_{\rm{D}}$} &\multirow{3}{8em}{$g=0.7$ $M_V=1028\en{GeV}$ $T_c=336\en{GeV}$} & $\alpha$ & 1497  & 4597 & $S_3=2136$ GeV \\ 
                & & $\beta/H$ & 36.8  & 49.5 & $\Delta V=2.75\cdot 10^9$ GeV \\ 
                & & $T_{\rm{n}}$ & 15.3 GeV & 11.4 GeV & $v=1489$ GeV \\ 
                \hline
                \multirow{3}{3.5em}{$SU(2)_{\rm{D}}$} &\multirow{3}{8em}{$g=2.0$ $M_V=7530\en{GeV}$ $T_c=2345\en{GeV}$} & $\alpha$ & 0.16  & 0.22 & $S_3=2.00\cdot 10^5$ GeV \\ 
                & & $\beta/H$ & 289  & 301 & $\Delta V=3.38\cdot 10^13$ GeV \\ 
                & & $T_{\rm{n}}$ & 1430 GeV & 1446 GeV & $v=7515$ GeV \\ 
                \hline
            \end{tabular}.    
        \end{center}
        \caption{Table with benchmark points used for the discussion of gravitaitonal wave signals. The two first benchmark points are from the $U(1)_{\rm{D}}$ model and the last is from the $SU(2)_{\rm{D}}$ model.}
        \label{tab:GWpoints}
    \end{figure}

    Notice that the first point is below one TeV, the trend we observed regarding the performance of the two simulation tools is noticeable, and the \texttt{BubbleProfiler} nucleation temperature is significantly below the value obtained from \texttt{CosmoTransitions}, affecting also the $\alpha$ and $\beta$ parameters. At this point, the critical temperature is $T_{\rm{c}} = 303$.
    
    In summary, both \texttt{CosmoTransitions} and \texttt{BubbleProfiler} show similar behavior for both models and are in reasonable agreement. For high masses the latter tool yields slightly higher nucleation temperatures and therefore $\alpha$ is also a bit lower and $\beta$ as indicated by Equation~\ref{eq:gwparams}.

    Having computed the relevant parameters for calculating GW spectra, we can consider the following equation for computing the total signal,
	\begin{gather}
		\Omega_{\rm{tot}}h^2 = 	\Omega_{\rm{col}}h^2  + \Omega_{\rm{sw}}h^2  + 	\Omega_{\rm{turb}}h^2,
	\end{gather}
	where the first term is the collision term, the second term is the sound wave term, and the last term is the turbulence term. The collisions from the bubbles themselves contribute to the GW spectra, but this term is more than 10 orders of magnitude lower than the other terms and plays no significant role in the result. The collisions also produce bulk motion in the fluid. This causes sound waves that result in the primary contribution to the GW spectra. Finally, there is also some turbulence caused by the collisions which contribute to the GW spectra \cite{mohamadnejad2020,alanne2020,kamionkowski1994}. The relevant equations for computing the GW spectra are \cite{ellis2020},
	\begin{gather}
		\Omega_{\rm{col}}h^2(f) = 0.5\cdot 10^{-5} v_w^2 \frac{H^2}{\beta^2} \paa{\frac{g_{e}}{100}}^{-\frac{1}{3}}\paa{\frac{\kappa_{\rm{col}}\alpha}{1+\alpha}}^2 \paa{\frac{f_*}{f_{\rm{col}}}}^3 \paa{1+2\paa{\frac{f_*}{f_{\rm{col}}}}^{2.07}}^{-2.18}, \\
		\Omega_{\rm{sw}}h^2(f) = 4.175 \cdot 10^{-6} v_w (H_* \tau_{\rm{sw}}) \frac{H}{\beta} \paa{\frac{g_{e}}{100}}^{-\frac{1}{3}} \paa{\frac{\kappa_{\rm{sw}}\alpha}{1+\alpha}}^2 \paa{\frac{f_*}{f_{\rm{sw}}}}^3 \paa{1+\frac{3}{4}\paa{\frac{f_*}{f_{\rm{col}}}}^{2}}^{-\frac{7}{2}},\\
		\Omega_{\rm{turb}}h^2(f) = 3.32 \cdot 10^{-4} v_w (1-H_* \tau_{\rm{sw}}) \frac{H}{\beta} \paa{\frac{g_{e}}{100}}^{-\frac{1}{3}}  \paa{\frac{\kappa_{\rm{sw}}}{1+\alpha}}^{\frac{3}{2}} \paa{\frac{f_*}{f_{\rm{turb}}}}^3\frac{\paa{1+\frac{f_*}{f_{\rm{turb}}}}^{-\frac{11}{3}}}{1+8 \pi f_*},
	\end{gather}
    for the collision, sound wave and turbulence terms respectively. Note we use $g_e \approx 103$. These terms are all corrected for redshifting, but the frequency $f_*$ is not redshifted to today. The different frequency terms are given as
    \begin{gather}
        f_{\rm{col}} =  \frac{1.1}{v_w} \frac{\beta}{H}, \quad f_{\rm{sw}} = \frac{1.16}{v_w}\frac{\beta}{H}, \quad f_{\rm{turb}}  =  \frac{1.75}{v_w}\frac{\beta}{H}
    \end{gather}
	and the frequency $f$ redshifted to today is given as,
    \begin{gather}
        f =  1.65 \cdot 10^{-5} \en{Hz} \frac{T_{\rm{reh}}}{100\en{GeV}}\paa{\frac{g_{e}}{100}}^\frac{1}{6} f_*,
    \end{gather}
    and the reheating temperature is approximated as \cite{ellis2018}
    \begin{gather}
        T_{\rm{n}} = T_{\rm{n}} \paa{1+\alpha}^{\frac{1}{4}}.
    \end{gather}

    To obtain the values for $\kappa_{\rm{sw}}$ it is necessary to determine the wall velocity $v_w$. To this end one needs to determine the Jouguet velocity \cite{lewicki2022}. This can be done using
    \begin{gather}
        v_J = \frac{1}{\sqrt{3}}\frac{1+\sqrt{3\alpha^2 + 3\alpha}}{1+\alpha}.
    \end{gather}
    From this one can compute the wall velocity
    \begin{gather}
        v_w = 
        \begin{cases}
        \sqrt{\frac{\Delta V}{\alpha \rho}} & \en{if}  \quad  \sqrt{\frac{\Delta V}{\alpha \rho}} < v_J,\\
        1 & \en{if} \quad \sqrt{\frac{\Delta V}{\alpha \rho}} \ge v_J,
        \end{cases}
    \end{gather}
    which in the case for this model based on the  Coleman-Weinberg mechanism yields $v_w=1$. Having investigated this value, we can use the appropriate equation for $\kappa_{\rm{sw}}$ \cite{espinosa2010, abe2023, lewicki2023}. In the limit of $v_w\rightarrow 1$, we use \cite{ellis2020}
    \begin{gather}
        \eval{\kappa_{\rm{sw}}}_{v_w \rightarrow 1} = \frac{\alpha_{\rm{eff}}}{\alpha}\frac{\alpha_{\rm{eff}}}{0.73+0.083\sqrt{\alpha_{\rm{eff}}}+ \alpha_{\rm{eff}}},
    \end{gather}
    where
    \begin{gather}
        \alpha_{\rm{eff}} = \alpha(1-\kappa_{\rm{col}}), \quad \kappa_{\rm{col}} \approx \frac{3}{2} \frac{\gamma_{\rm{eq}}}{\gamma_*}, \\
        \gamma_{\rm{eq}} = \frac{\alpha- \alpha_{\infty}}{\alpha_{\rm{eq}}}, \quad  \alpha_\infty = \frac{15g^2}{\pi^4 g_e}\paa{\frac{v}{T_{\rm{reh}}}}^2, \quad \alpha_{\rm{eq}} = \frac{180g^3}{\pi^3 g_e} \frac{v}{T_{\rm{reh}}},\\
        \gamma_* = \frac{2}{3} \frac{R_*}{R_0}, \quad R_* = \paa{8\pi}^{\frac{1}{3}}\frac{v_w}{\beta}, \quad R_0= \paa{\frac{3 S}{2\pi \Delta V}}^{\frac{1}{3}}.
    \end{gather}
    The sound wave duration is defined as
    \begin{gather}
        H_* \tau_{\rm{sw}} \equiv \min \paa{1, H_* \tau_{\rm{sh}}}, \quad H_* \tau_{\rm{sh}} \approx \frac{\paa{8 \pi}^{\frac{1}{3}} \max \paa{v_w, c_s}}{\bar{U}_f} \frac{H}{\beta}, \quad \bar{U}_f \simeq \sqrt{\frac{3}{4}\frac{\kappa_{\rm{sw}} \alpha}{1 + \alpha}},
    \end{gather}
    where $c_s = \sqrt{1/3}$. Since $v_w \rightarrow 1$, this reduces the shock duration to
    \begin{gather}
        H_* \tau_{\rm{sh}} \approx \frac{\paa{8 \pi}^{\frac{1}{3}}}{\bar{U}_f} \frac{H}{\beta}
    \end{gather}
    
    \begin{figure}
        \centering
        \includegraphics[height=10cm]{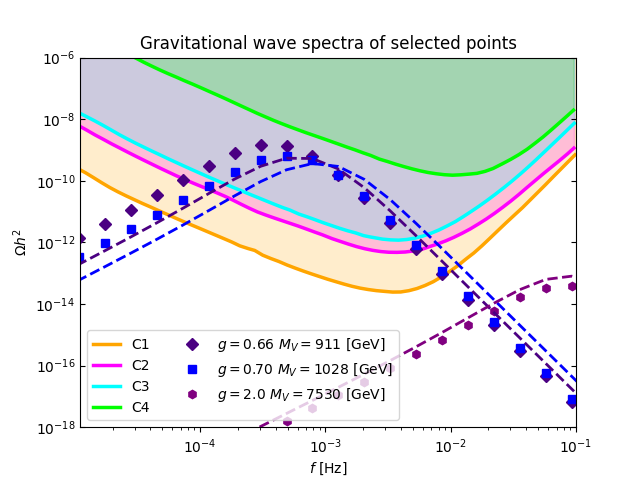}
        \caption{The GW spectra for two different sets of transition parameters for the $U(1)_{\rm{D}}$ model and one for the $SU(2)_{\rm{D}}$ model ($g=2.0$, $M_V=7530$) computed with \texttt{CosmoTransitions} (dashed lines) and \texttt{BubbleProfiler} (dotted lines). The sensitivity curves (C1-C4) of the LISA detector are also shown \cite{lisa2016}. According to this result, the signals from this model are strong enough for LISA to detect the GW signal from the phase transition.}
        \label{fig:waves}
    \end{figure}
    
    The result of this computation can be seen in Figure \ref{fig:waves} for the three benchmark points and their respective parameters, two from the $U(1)_{\rm{D}}$ model and one from the $SU(2)_{\rm{D}}$. The dominant terms are the turbulence, $\Omega_{\rm{turb}}$, and the sound wave $\Omega_{\rm{sw}}$ even though the duration of the sound wave is rather short: $H_* \tau_{\rm{sw}} < 0.15$.
    
    The marker shape indicates the parameter as shown in Figure \ref{fig:constraintCombined}. The diamond and square shapes are from the $U(1)_{\rm{D}}$ model. For the $SU(2)_{\rm{D}}$ model we have the high mass case marked by the hexagon shape. The projected sensitivity curves (for the configurations C1-C4) for the LISA detector are also shown \cite{lisa2016}, and one can see that for the $U(1)_{\rm{D}}$ model the signal should be detectable by three out of four configurations at a frequency at around 1 mHz, but for the $SU(2)_{\rm{D}}$ model the mass becomes too high and we need other future experiments to detect such high DM mass models such as the proposed TianQin detector \cite{luo2016}.
	
\section{Discussion and conclusions} 
    We have investigated two vector DM  models in light of existing DM direct detection experiments and future GW experiments .Both of the models investigated in this work are already strongly constrained by direct detection. For the $SU(2)_{\rm{D}}$ model, this is, in particular, due to recent results from the LZ experiment which has ruled out most of parameter space consistent with a full relic abundance from freeze-out in the range $M_V\in (1-8)$ TeV and with XENONnT either the DM will be detected or the entire parameter space above the neutrino floor will be ruled out  
    as shown in Figure \ref{fig:constraintCombined}.
    
    GW signals in both models have been discussed in earlier literature. In our analysis we find that 
    results differ significantly between different numerical implementations.
    Recently, the $SU(2)_{\rm{D}}$ model was discussed in \cite{kirkla2022}, and we find that their results for the $\alpha$ and $T_{\rm{n}}$ parameters agree with our findings. 
    
    Regarding the $U(1)_{\rm{D}}$ model, it was previously suggested that GW signals could be used to probe the model in case the direct detection cross section remains below the neutrino floor \cite{mohamadnejad2020}. We agree with this conclusion, but numerically we find differences to \cite{mohamadnejad2020} in the GW parameters. While we can reproduce the critical temperature reported, the nucleation temperature and the $\alpha$ and $\beta$ parameters differ from those reported in \cite{mohamadnejad2020}. Their results were obtained with the \texttt{AnyBubble} package \cite{masoumi2017}, for which we failed to obtain results in agreement with the other two numerical implementations used in this work.
    
    This raises the question of comparability between the phase transition parameters obtained via the various numerical implementations available. This issue has been investigated in \cite{athron2019}, where a fairly good agreement between \texttt{BubbleProfiler} and \texttt{CosmoTransitions} is observed. This is compatible with our findings.
    
    The finite temperature potential in both cases leads to a strong first-order electroweak phase transition. The $U(1)_{\rm{D}}$ model can produce significant GW signals, which can be detected by LISA \cite{lisa2016} and future experiments would be able to test the $SU(2)_{\rm{D}}$ model also in the high DM mass regime.  

    In conclusion, in this paper we have made two main observations. First, the $U(1)_{\rm{D}}$ model is ruled out by direct detection for all DM masses in the range $M_V\in (1-4)$ TeV, except for a small gap near the SM-Higgs and scalon interference. Extrapolation from Figure \ref{fig:constraintsU1} suggests that  higher masses are also excluded, though it is beyond this paper to look into very heavy DM masses beyond 100 TeV where the direct detection constraints will begin to relax. This means that, except masses in the interference zone and possibly very heavy masses, this model is no longer viable. And with upcoming experiments this interference gap is closing. Given a suitable DM signal is found around in the interference zone, then one can see from Figure \ref{fig:waves}, that one should find a GW signal with the upcoming LISA experiment \cite{lisa2016}.

    Second, for the $SU(2)_{\rm{D}}$ model we find that for DM masses beyond 8 TeV the model is still viable, but the heavy mass reduces the possible GW signal and LISA is not expected to see a signal from this model, but other future experiments with higher precision might detect it. Conclusively, the parameter space of vector DM Higgs portal models is dramatically shrinking and the models may become excluded in the near future.

\acknowledgements{
The financial support from Academy of Finland, project $\# 342777$, is gratefully acknowledged. MTF and MR acknowledge partial funding from The Independent Research Fund Denmark, grant numbers DFF 6108-00623 and DFF 1056-00027B, respectively. MET acknowledges funding from Augustinus Fonden, application $\#22-19584$, to cover part of the expenses associated with visiting  the University of Helsinki for half a year. We thank Marek Lewicki for pointing out an improved calculation procedure for GW signals in CW-like theories.}

\appendix

\section{Supercooling, inflation and reheating}\label{appA}
    The investigation of the GW spectra leads to the discussion of supercooling in the models presented. As shown in the GW section, there are orders of magnitude in the difference between the critical and nucleation temperatures at the low mass scale. As discussed in the other papers, this can lead to different kinds of phenomena including inflation, filtering, and reheating \cite{hambye2018, Baldes:2021aph, baker2020}. These effects are expected to affect the GW signal for low masses, and it might affect some of the results even presented in Figure \ref{fig:waves}, but it is beyond this paper to look at the details of this. As discussed in a recent paper, the Universe could escape inflation via bubble nucleation or via quantum tunneling, two different scenarios leading to different GW signals \cite{ville2021}.
    
	We would, however, like to highlight the fact that strong supercooling from hundreds of GeV to the QCD scale
	might not be a big issue for the models. The bigger the	supercooling, the greater the inflation, as the scalon Higgs
	field will be stuck in a false vacuum acting like a	cosmological constant. The main inflationary constraint
	is that any amount of inflation resulting from supercooling should not exceed the max number of $e$-folds:
    \begin{gather}\label{eq:maxefold}
        N_{\rm{max}} = 23.8 + \ln\frac{T_{\rm{R}}}{\rm{TeV}},
    \end{gather}
    where $T_{\rm{R}}$ is the reheating temperature after the inflationary epoch and one finds that this limits the temperature to $T_R < 6.6\cdot 10^{15}$ GeV \cite{ville2021}. To compute the reheating temperature, we are interested in computing the decay of the inflaton-like field which in this case is the scalon Higgs field $S$ for the $U(1)_{\rm{D}}$ model. Because of the mass constraints, only the scalon Higgs is kinetically allowed to decay as $\Gamma(h_S \rightarrow h,h)$, but this requires a DM mass of $M_V > 1$ TeV. From the Lagrangian, we find that the Feynman rule for this vertex and this yields the decay
	\begin{gather}
		\Gamma(h_S \rightarrow 2h) = \frac{\sqrt{M_S^2 - 4 M_h^2}}{32 \pi M_S^2} \abs{\mathcal{M}}^2,
	\end{gather}
	where
	\begin{gather}
		 \abs{\mathcal{M}}^2 = \paa{\frac{M_h^2}{4 v_1}\paa{5 + 3\cos(4 \alpha)}\sin(\alpha)}^2.
	\end{gather}
    We can furthermore include decays into quarks and leptons:
	\begin{gather}
		\Gamma(h_S \rightarrow f\bar{f}/\ell\bar{\ell}) = \frac{N_C }{8\pi}\frac{m_b^2}{v_1^2}M_S\sqrt{1-\frac{4m_b^2}{M_\phi^2}}\sin(\alpha)^2,
	\end{gather}
	where $N_C = 3$ for fermions and $N_C = 1$ for leptons. Using the decays one can calculate the reheating temperature $T_R$ using the following equation \cite{schmitz2010}
	\begin{gather}
		T_{\rm{R}} \approx 0.2 \paa{\frac{200}{g_*}}^{1/4}\sqrt{\Gamma M_{\rm{pl}}},
	\end{gather}
	where $M_{\rm{pl}}$ is the reduced Planck mass and $g_* =103$. Considering a rather low mixing value $0 \le \alpha \le \frac{\pi}{64}$ and a mass range of $250$ GeV $\le M_V \le 2500$ GeV the reheating temperature is somewhere around 0.1-1.6 PeV, yielding a mass of the scalon field around $1$ GeV $< M_S < 200$ GeV. This is so hot that the Universe will reheat back to a temperature much hotter than the scales of freeze-out. It also satisfies the constraint from Equation \eqref{eq:maxefold}; thus, it is not too hot and not causing too much inflation. One can repeat this exercise for the $SU(2)_{\rm{D}}$, but the result is roughly the same with the main differences being a slightly heavier scalon mass, $1$ GeV $< M_S < 350$ GeV, and higher reheating temperature 0.1-2 PeV. Conclusively, dark matter production can take place via freeze-out as the Universe subsequently cools down again.
 
\section{Model implementation in \texttt{CosmoTransitions}}\label{appB}
    For the implementation of the model in \texttt{CosmoTransitions} we feed in the tree-level potentials as shown in Equation \eqref{eq:potu1} and \eqref{eq:potsu2}. Then we manually implement the field-dependent mass matrix, indicated by the tilde, with the massive SM bosons, plus the new bosons, and the top quark,
    \begin{gather}
        \Tilde{M}_W^2 = \frac{g_W^2}{4}h_1^2, \quad \Tilde{M}_Z^2= \frac{g_W^2 + g_Z^2}{4}h_1^2, \quad \Tilde{m}_t^2=\frac{\lambda_t^2}{2}h_1^2,
    \end{gather}
    where the Yukawa coupling of the top quark is $\lambda_t =1$. The DM candidates have their respective implementations for each model where
    \begin{gather}
        \Tilde{M}_V^2 = c_Vg^2 h_2^2, 
    \end{gather}
    with $c_V=1\,(1/4)$ for the U(1)$_D$ (SU(2)$_D$) model,
    and then the scalar mass matrices yield
    \begin{gather}
    	\Tilde{M}_{h,S}^2 = \frac{1}{4} \Bigg[h_1^2 \paa{\lambda_h +  2\lambda_{\phi h}} + h_2^2(\lambda_\phi + 2\lambda_{\phi h}) \nonumber \\
		\pm \sqrt{h_1^4 \paa{\lambda_h - 2 \lambda_{\phi h}}^2 +h_2^4 \paa{\lambda_\phi  -2\lambda_{\phi h}}^2 + 2h_1^2 h_2^2 \paa{2\lambda_h \lambda_{\phi h} + 28 \lambda_{\phi h}^2 - \lambda_h \lambda_\phi + 2\lambda_\phi \lambda_{\phi h}}} \Bigg].
    \end{gather}
    The mass parameters are expressed in terms of the fields, as this is how they are implemented in the code since we are considering the thermal evolution where the minimal field value is not necessary that of the VEVs. At zero temperature is, however, where we have the minima such that $h_1 \rightarrow v_1$ and $h_2 \rightarrow v_2$.
    
    A custom solution is made for computing the beta value. This is done by simply calculating the action divided by the temperature around the point of the nucleation temperature, making a fit to those plots, and taking the derivative, etc. Some tweaks have been done to the source code to make this work and also to improve the precision at low nucleation temperatures.

\section{Model implementation in \texttt{BubbleProfiler}}\label{appC}
    For this package, we give it the full thermal potential in Equation \eqref{eq:firstorder}, but instead of evaluating the thermal integral an approximation is made using Bessel function\cite{mohamadnejad2020}. Specifically, we use the modified Bessel functions $K_2(kx)$ as follows
    \begin{gather}
        \int_{0}^{\infty} y^2 \ln{\paa{1 \mp e^{-\sqrt{y^2 \pm x^2}}}}\dd y = - \sum_{k=1}^3 \frac{x^2}{k^2} K_2(kx) - \sum_{k=1}^2 \frac{(-1)^kx^2}{k^2} K_2(kx),
    \end{gather}
    where $x=M_S/T$ and $k=3(2)$ for bosons(fermions) is sufficient for all practical purposes as also discussed in \cite{mohamadnejad2020}. In the case of \texttt{CosmoTransitions} we kept the default of $k=8$, but according to tests this does not bring any significant numerical improvement.
    
    The \texttt{BubbleProfiler} is written in C++, but comes with a command line interface (CLI). Using this one can implement simple potentials like polynomials. In order to avoid implementing all these functions, we created a python interface where we implement the model in python. Then we create a higher-order polynomial fit to the full potential. This polynomial is then fed to \texttt{BubbleProfiler} via the CLI together with other relevant parameters. We compute several points around the nucleation temperature and make a fit to that, from there we determine the $\beta$ and $T_{\rm{n}}$ value, and the latter is then used to find $\alpha$.

\section{Computing parameters of the EWPT}\label{appD}
\begin{figure}
        \centering
            \subfigure[Using \texttt{BubbleProfiler} for computing parameters.]{
                \label{fig:bubblebeta}
                \includegraphics[height=6.4cm]{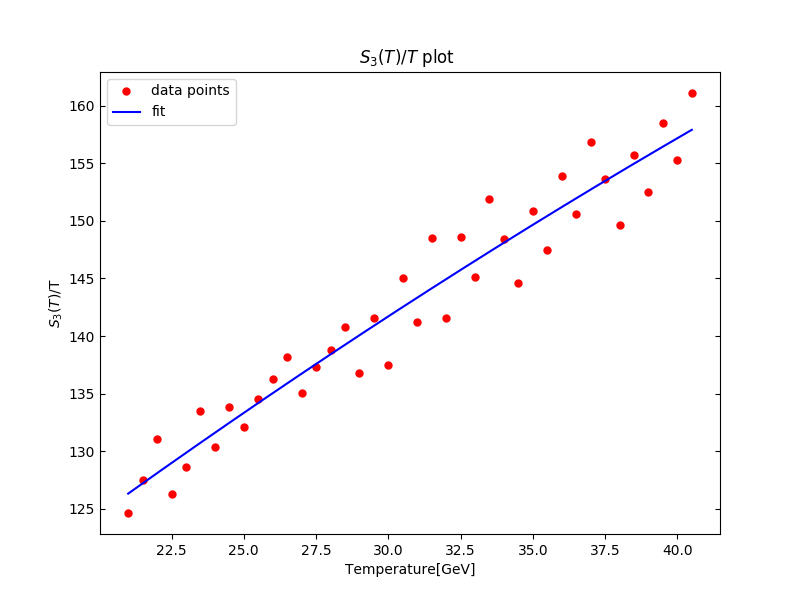}}
            \subfigure[Using \texttt{CosmoTransitions} for computing parameters.]{
                \label{fig:cosmobeta}
                \includegraphics[height=6.4cm]{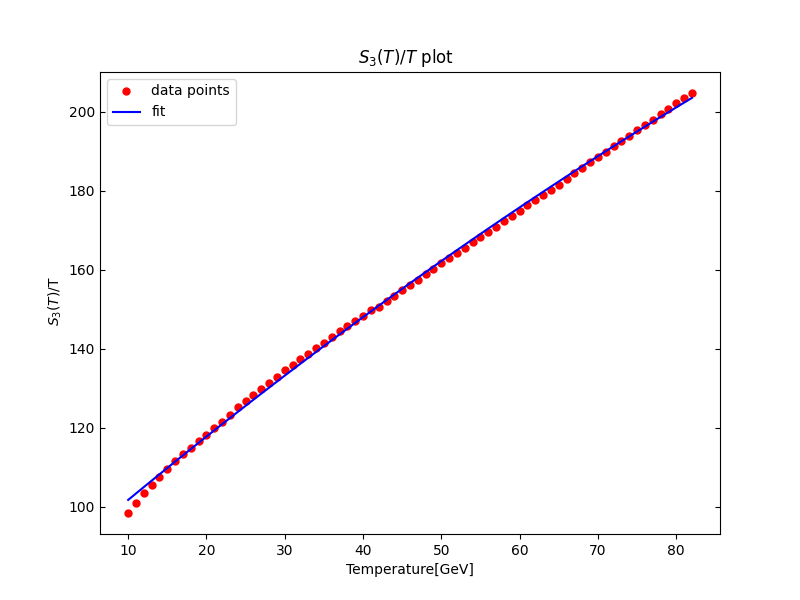}}
        \caption{A comparison of the apparent error when computing the $\beta$ and $T_{\rm{n}}$ parameters when computing the EWPT parameters in $U(1)_{\rm{D}}$ model for $g=0.75$ and $M_V=1184$. Note that the temperature range is different for the implementation; thus, the range is different in the plots.}
        \label{fig:beta}
    \end{figure}

    The essential computation for the phase transition is finding the relationship between the action and temperature. The nucleation temperature condition is defined as
    \begin{gather}
        \eval{\frac{S(T)}{T}}_{T_{\rm{n}}} \approx 140,
    \end{gather}
    thus when the action divided by the temperature is equal to 140. We can compute the action and by dividing by the temperature a plot of this relationship can be obtained as seen in Figure \ref{fig:beta}.
    
    Given some data points, it is possible to make a fit, and from that read off the $T_{\rm{n}}$ value. Furthermore, the fit is also a function of $S_3(T)/T$, which can, thus, be used to compute $\beta$. Now recall that $\alpha$ is evaluated at the nucleation temperature and $\alpha \propto 1/T_{\rm{n}}^4$, thus, the value of $\alpha$ is also highly dependent on the nucleation temperature. Since our \texttt{BubbleProfiler} result, in general, yields a slightly higher nucleation temperature we get a lower value of $\alpha$ as discussed in the GW section. For lower masses, the \texttt{BubbleProfiler} result yield significantly lower nucleation temperatures suggesting, that our implementation might not be as good in this regime.
    
    Looking at Figure \ref{fig:beta}, we see that the apparent error of the \texttt{BubbleProfiler} is significantly higher than the error from the \texttt{CosmoTransitions} result. This may be attributed to the fact that we used an approximated potential via our custom Python interface instead of implementing the model using C++. This leads us to consider the \texttt{CosmoTransitions} as the better result in this paper even though \texttt{BubbleProfiler} is claimed to be more accurate \cite{athron2019}.
    
%%%%%%%%%%%%%%%%%%%%%%%%%%%%%%%%%%%%%%%%%%%%%%%
% Bibstyle

\bibliographystyle{hunsrtnat}

%%%%%%%%%%%%%%%%%%%%%%%%%%%%%%%%%%%%%%%%%%%%%%%
% Bibliography

\bibliography{main}

\end{document}